\documentclass{evn2004}
\usepackage{txfonts}
\usepackage{graphicx}
\begin{document}
\setcounter{page}{231}

   \title{Continuum EVN and MERLIN observations of Ultra Luminous Infrared Galaxies}

   \author{A.G. Polatidis \inst{1}
          \and
J.E. Conway\inst{2}
          }

   \institute{Max Planck Institut f\"{u}r Radioastronomie, Bonn, Germany
         \and
Onsala Space Observatory, Sweden}

   \abstract{Radio imaging of ULIR galaxies is ideal to explore the connection
between the starburst and the AGN phenomenon since it is unaffected
by dust obscuration, and provides the required high angular resolution
to distinguish between an AGN and starburst emission. We have made
combined 18cm radio continuum, EVN and MERLIN observations of 13
ULIRGs that have the parsec and deci-parsec scale
resolution necessary to distinguish between an AGN and supernovae
remnants at the centres of these galaxies, and assess the contribution
of each to the total energy distribution. Images of four galaxies are
presented here.
   }

   \maketitle
%

\section{Introduction}

To date the best studied sample at high resolution of ULIRG galaxies
is the IRAS Bright Galaxy sample (hereafter the BGS sample). Condon et
al. (\cite{condon90};\cite{condon91}) observed over 40 compact sources
with the VLA from 1.6 to 8 GHz. Lonsdale et al. (\cite{L93}, L93) made
snapshot VLBI images of 31 sources from Condon et
al. (\cite{condon91}); of these 21 were detected in VLBI baselines
greater than 10 M$\lambda$, prompting Londsale et al. to suggest that
an AGN was responsible for the VLBI structure. However, Smith et
al. (\cite{Smith98a}), in full track observations of Arp 220 showed
that the VLBI structure consists of radio supernovae or supernovae
remnants. Following this, Smith et al. (\cite{Smith98b}) modeled the
L93 data fitting clustered luminous compact radio supernovae, showing
that for most sources the VLBI structure could be explained by
starburst activity while for only a few sources the presence of an AGN
was invoked to account for the parsec scale structure.

We are engaged in a systematic survey of the complete sample of the 19
northern ($\delta>0$) compact ULIRGs in the BGS sample which showed
evidence for detections $>$1mJy on at least one baseline longer than
10 M$\lambda$. Of these 19 compact ULIRGs six (IIIZw 35, Mrk 231, Mrk
273, Arp 220, NGC 7469 and Mrk 331) were already well observed at the
same resolution, hence we were left with 13 objects to observe.

We have observed the sources with a combined MERLIN and EVN array at
18cm so that we can separate the diffuse emission at MERLIN scales
which can be due to thermal emission combined with synchrotron
emission from electrons from supernovae diffused into the ISM, from
the compact radio emission which can be due to an AGN or to
synchrotron emission associated with single or clustered radio
supernovae or supernovae remnants (SNRs).


\section{Observations}

Initially the five brighter sources were observed with the combined
MERLIN and EVN array, in February 2002. The remaining 8 sources were
observed in May 2003. In this paper we present preliminary
results from the February 2002 observations.

The sources were observed with six telescopes of the EVN (Torun,
Medicina, Onsala, Effelsberg, Westerbork, and the Jodrell Bank Lovell
and Mark 2 telescopes alternating during the observations) with a
sampling rate of 256 Mbits/s over a 32 MHz bandwidth. The MERLIN
observations had a 15GHz bandwidth in overlapping frequency range.
Due to the relative weakness of the sources, the observations were
made in phase-referencing mode with a cycle of 7 min on the source and
3 minutes on the nearby phase calibrator.

The observations were partially successful. Of particular importance
was the failure of the Cambridge telescope which would have provided
the common baseline (Lovell - Cambridge) to facilitate the
cross-calibration of the two arrays, and the subsequent reduction of
the MERLIN resolution to $\sim$ 250 mas.  Furthermore the loss of the
Lovell telescope for 40\% of the run, although it was replaced by the
Jodrell Bank Mark 2 antenna, resulted in reduction of sensitivity of
the array. Consequently the array consisted of the 6 MERLIN telescopes
and 6 EVN antennae at the most for each source.

\section{Results}

The EVN data were correlated at the JIVE data processor and were then
processed with the AIPS software suite. Initial amplitude calibration
and fringe fitting was done with the EVN pipeline and was refined
later.

\subsection{UGC 05101}

   \begin{figure*}
   \centering
   \includegraphics[width=7cm]{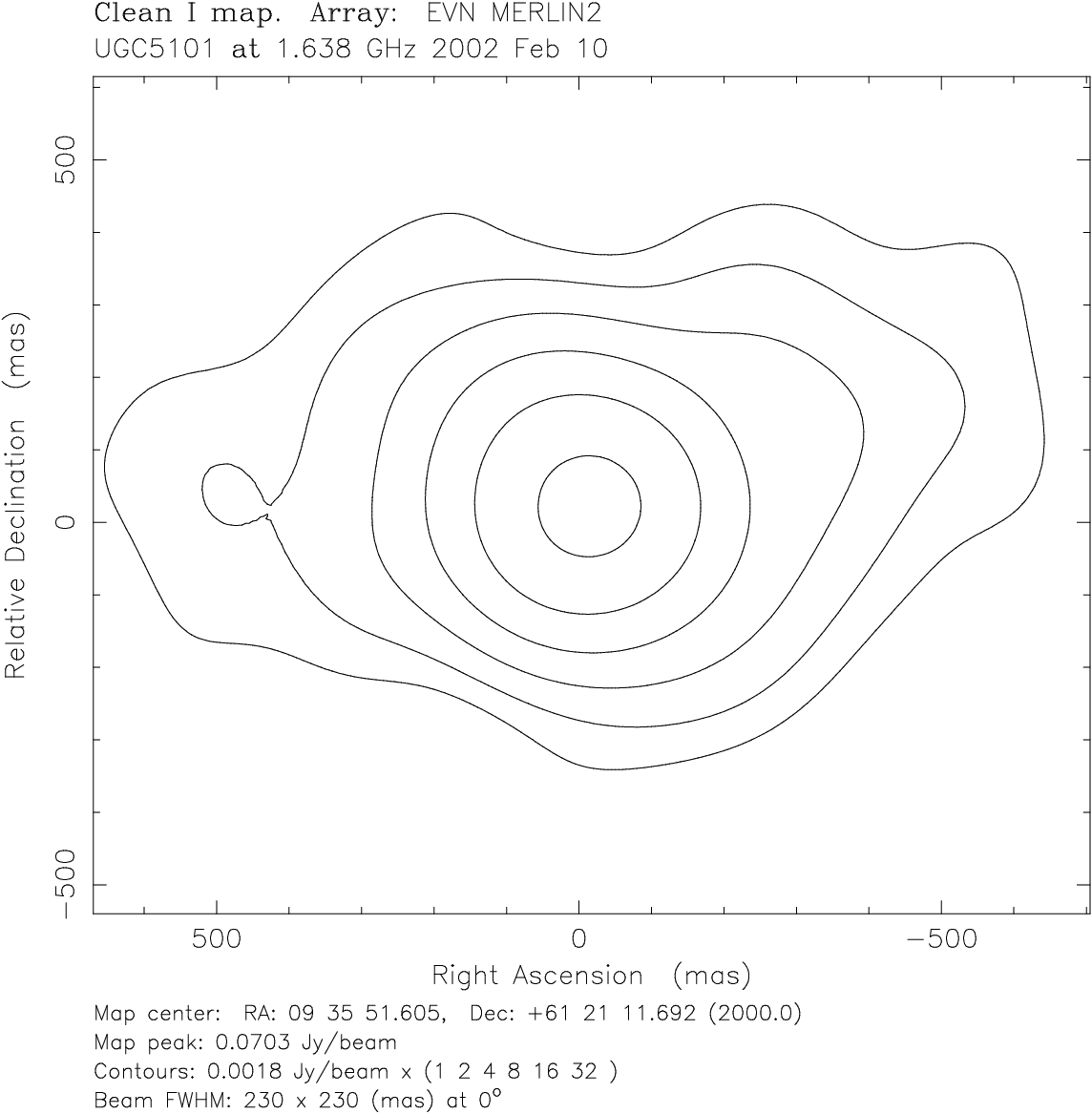} 
\includegraphics[width=7cm]{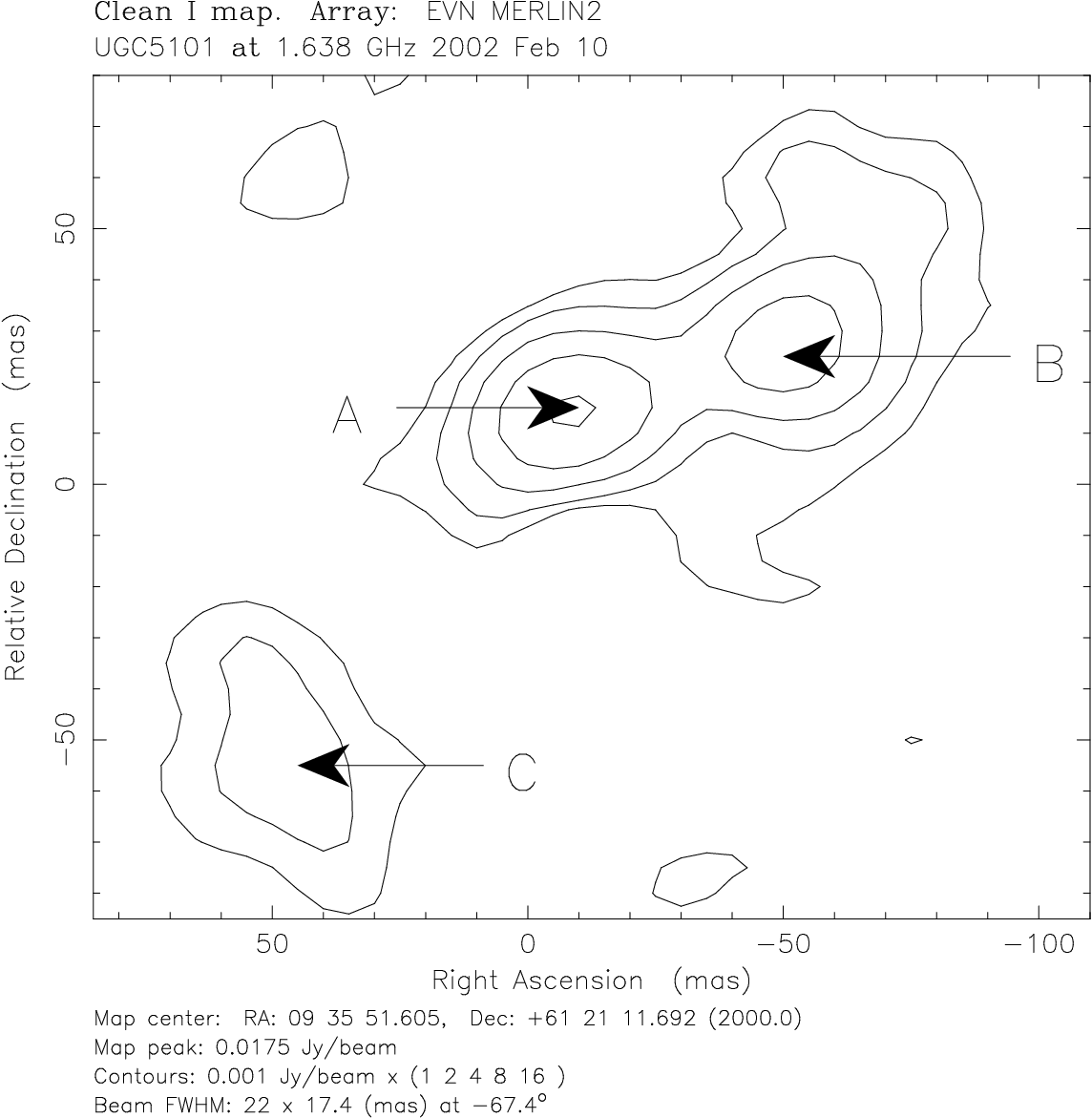}
   \caption{Left: The MERLIN image of UGC 05101 with a restoring beam of
  230 mas. The peak is 70.3 mJy/bm and the contours are drawn at
  1,2,4,8,16,32 x 1.8 mJy/bm. Right: The EVN image of UGC 05101
  restored with a beam of 22x17.4 mas.  The
  peak is 17.5 mJy/bm and the contours are drawn at 1,2,4,8,16 x 1mJy/bm.
            \label{fig5101}
           }
    \end{figure*}

UGC 05101 at VLA scales shows a $\sim$1.57 $\times$1.14 arcsec (1110
$\times$ 810 pc, at a distance d=158 Mpc) disk--like emission
elongated roughly east--west, dominated by a central bright component
(seen clearly in Condon et al's (\cite{condon91}) 8.4 GHz image). At
the MERLIN resolution of 270 mas (Fig. \ref{fig5101}) we detect a
similar structure measuring 933$\times$533 pc (roughly 1.3 $\times$
0.75 arcsec). The flux density in the MERLIN image is consistent with
that of Condon et al. (\cite{condon90}) and Thean et
al. (\cite{thean01}).

In the highest resolution (15 pc) EVN image $\sim$53\% of the flux
density is recovered in an apparently S-shaped source extending
roughly 128 $\times$140 parsec. Three major components (A-C),
relatively compact (5-13 pc) are embedded in weak extended radio
emission, visible in the lower resolution images that are not
presented here. Component A is closer to the peak brightness of the
MERLIN image. The brightness temperatures of components A and B are
$\sim$ 10$^{7}$K while that of the more resolved component C is 7
10$^{6}$K, while their luminosities range from 5 10$^{39}$ to
10$^{40}$ erg/s.

Lonsdale et al. (\cite{L03}) with higher resolution 5 GHz global VLBI
observations find a consistent nuclear structure, while they resolve A
and B in two compact subcomponents each.

These compact components are $\sim$50 times more luminous than the
brightest radio supernovae (RSN) in Arp220. They are also resolved and
their sizes are too large for single RSN. It appears that the nuclear
structure of UGC 05101 can not be associated with RSN but is rather
caused by an obscured AGN. This is consistent with Imanishi et al.'s
(\cite{imanishi03}) detection of Fe K$\alpha$ line emission in X-rays,
though the X-ray spectrum was measured with an aperture of 8 arcsec
(5.6 kpc) which covers the whole radio source and hence it is not easy
to orrelate the position of the X-ray source and the radio
emission. Multi-frequency observations are needed to further explore
this possibility.

\subsection{NGC 6286}

   \begin{figure*}
   \centering
   \includegraphics[width=6cm]{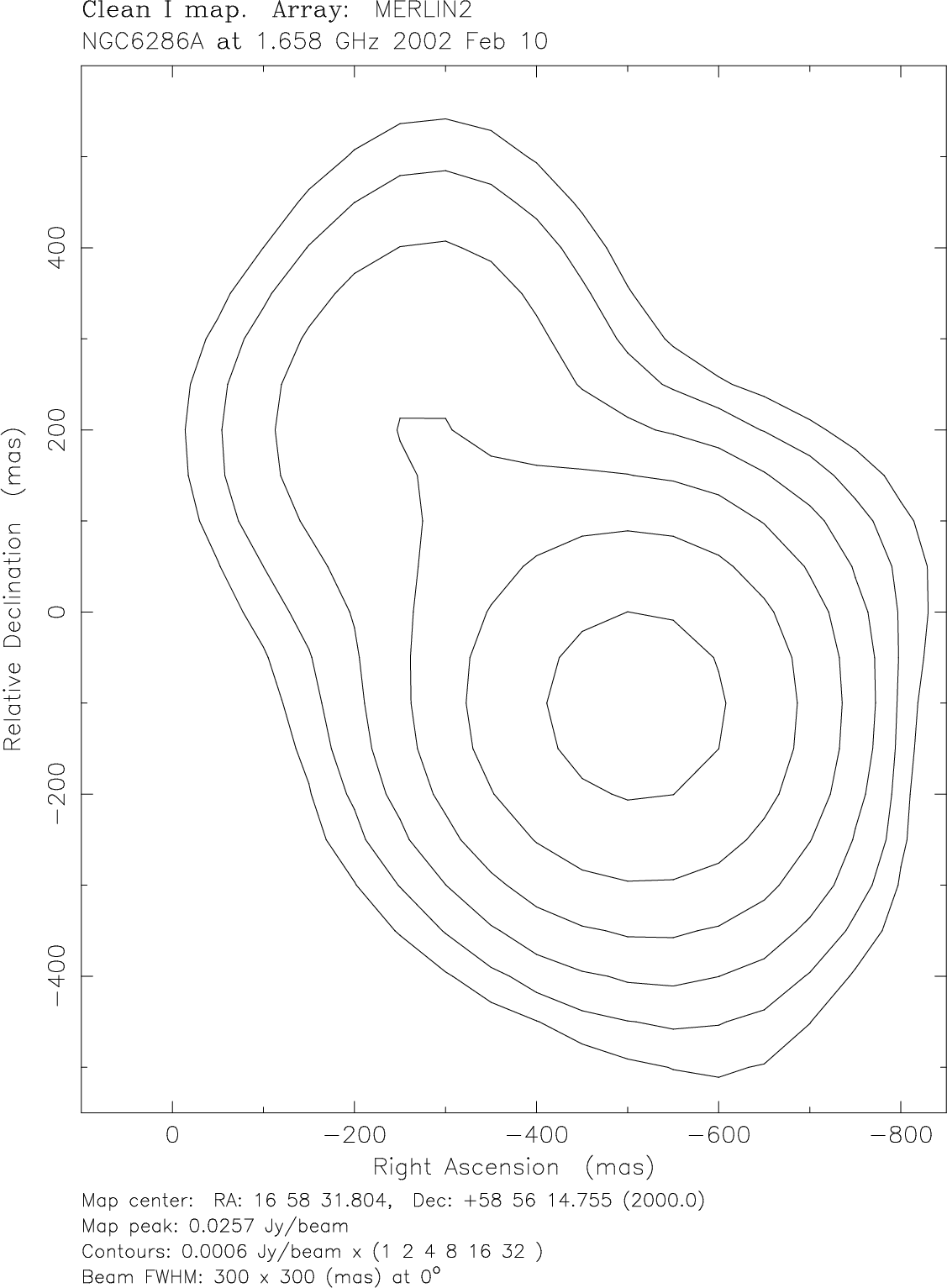} 
\includegraphics[width=6cm]{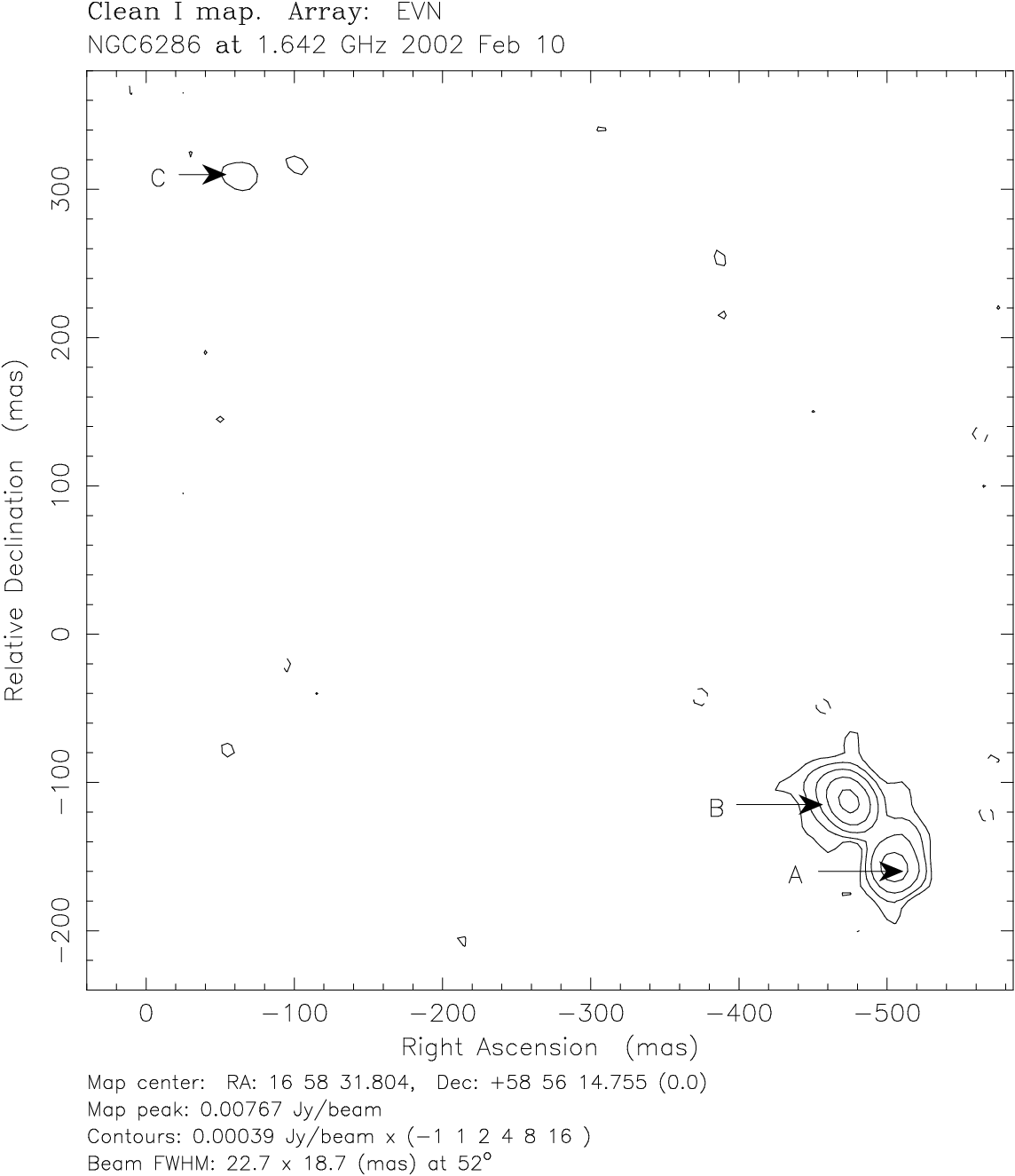}
   \includegraphics[width=6cm]{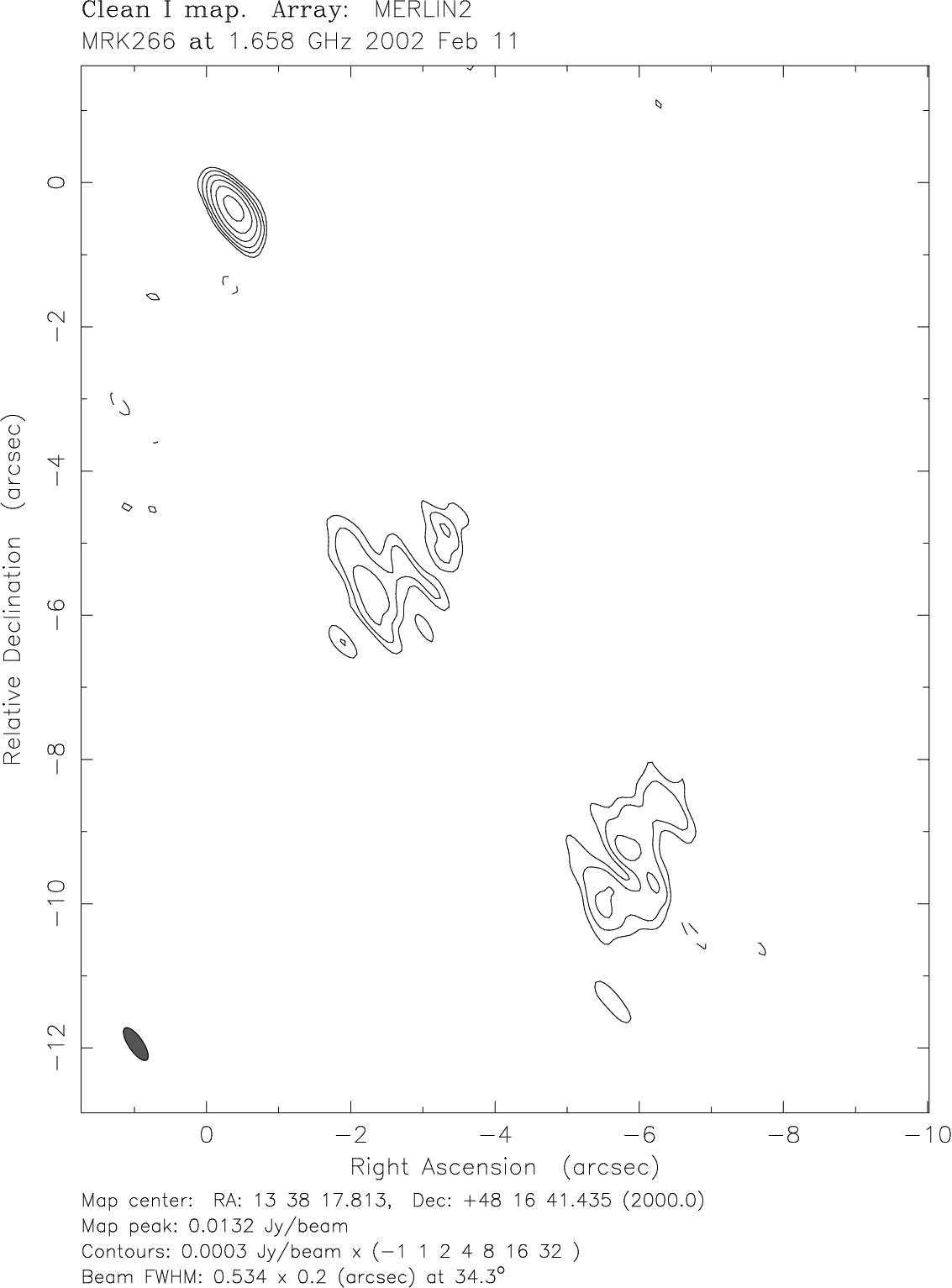} 
\includegraphics[width=6cm]{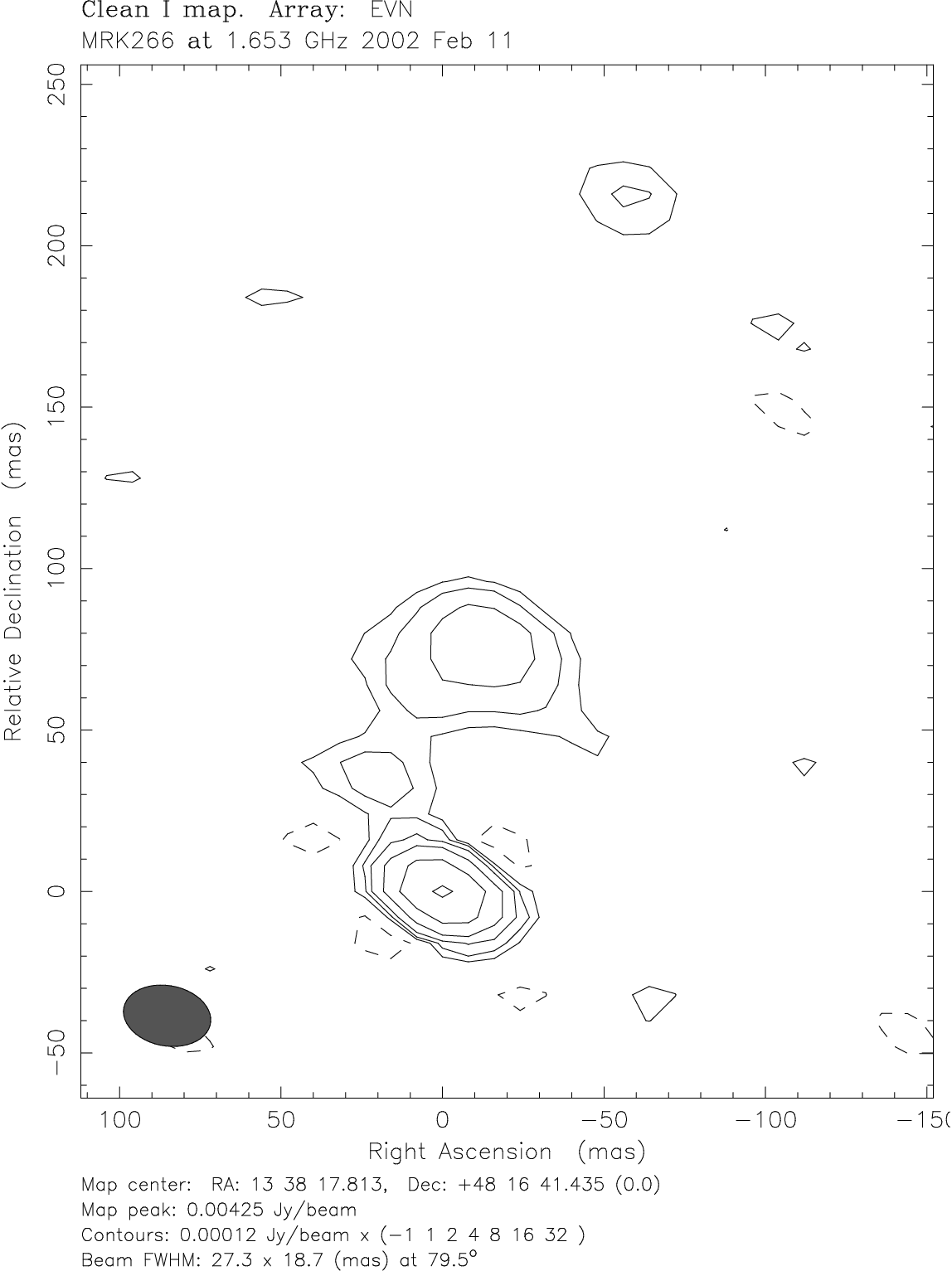}
   \caption{Top Left: The MERLIN image of NGC6286 with a restoring beam of
  300 mas. The peak is 25.7 mJy/bm and the contours are drawn at
  1,2,4,8,16,32 x 0.6 mJy/bm. Top Right: The EVN image of NGC6286 restored
  with a beam of 22.7x18.77 mas.  The peak is 7.6 mJy/bm and the
  contours are drawn at 1,2,4,8,16 x 0.39mJy/bm. Bottom Left: The MERLIN image of Mrk 266 with a restoring beam of
  534x200 mas. The peak is 13.2 mJy/bm and the contours are drawn at
  1,2,4,8,16,32 x 0.3 mJy/bm. Bottom right: The EVN image of Mrk 266
  restored with a beam of 27x18 mas.  The
  peak is 4.2 mJy/bm and the contours are drawn at 1,2,4,8,16 x 0.12mJy/bm.}
\label{fig6286}
\end{figure*}

NGC 6286 is an edge-on spiral galaxy (at a distance d=73 Mpc) in an
interacting pair with NGC 6285 (~1'.5 , 30 kpc) classified as a
LINER. Condon et al. (\cite{condon90}) interpret the 1.6 GHz VLA
image, showing a 20'' (6.8 kpc) resolved source elongated along the
optical axis as starburst induced.

Our MERLIN image (Fig. \ref{fig6286}) recovers 37 mJy ($\sim$ 25\% of the
VLA flux density) and consists of two-components separated by 450 mas
(150 pc) at the same orientation as the VLA image. At the 25 mas
resolution of the EVN image, the brighter MERLIN component is resolved
in two components (A, B) separated by 18.4 pc, with the brightest B
located at the northeast. 200 parsec northeast of B, we detect the
resolved out component C, which is the weaker component of the double
in the MERLIN image.

All three components in NGC6286 are resolved with sizes between 4 and
10 parsecs, brightness temperatures ranging from 1.02 10$^{6}$K (C) to
8.9 10$^{6}$K (B) and luminosities from 10$^{37}$ to 10$^{39}$ erg/s,
hence a few times higher than the RSN in Arp 220.

The fact that they are resolved argues against single RSN. It is
plausible that one of the compact components may be due to a clump of
supernovae. It is also possible that the brightest component are due
to AGN activity.

\subsection{Mrk 266}

Mrk 266 is a system of interacting galaxies (at a distance d=113 Mpc)
showing an interaction region in the middle and tidal tails. Mazzarella
et al. (\cite{ma88}) classify the northern as a LINER and the southern galaxy
as a Seyfert 2 type. VLA and MERLIN images show at least three
co-linear components of which the outer are associated with the two
interacting galaxies and the central may be a region of synchrotron
emission stimulated by the collision of the two galaxies with the
interaction region. With increasing resolution the two southern
components become resolved while the northern (which is the target of
our EVN observations) remains compact.

In our MERLIN image (Fig. \ref{fig6286}), the northern component has a
flux density of 21mJy, consistent with the image of Thean et al.
(\cite{thean01}) and a spectral index $\alpha_{1.6-8GHz}$=0.93. The
higher resolution EVN observations (beam=21mas, 107 pc) show a
relatively compact (2$\times$2pc) component (3.4 mJy) and arc-like
emission which culminates in a more extended (18$\times$14pc) weaker
(1.7 mJy) subcomponent $\sim$41pc, while a third weak component is
visible $\sim$100pc further north. Their brightness temperature range
from 2.3 10$^{7}$K to 3.3 10$^{5}$K and their luminosity from 4
10$^{38}$ to 8.2 10$^{38}$ erg/s.

Combining our EVN image with archival MERLIN 5GHz data, we find that
the brightest, compact component has a steep spectrum of
$\alpha_{5-1.6GHz}$=0.681, while weaker and more extended component
has a rising spectrum of $\alpha_{5-1.6GHz}$=-0.023.

\subsection{NGC 2623}

   \begin{figure*}
   \centering
\includegraphics[width=7cm]{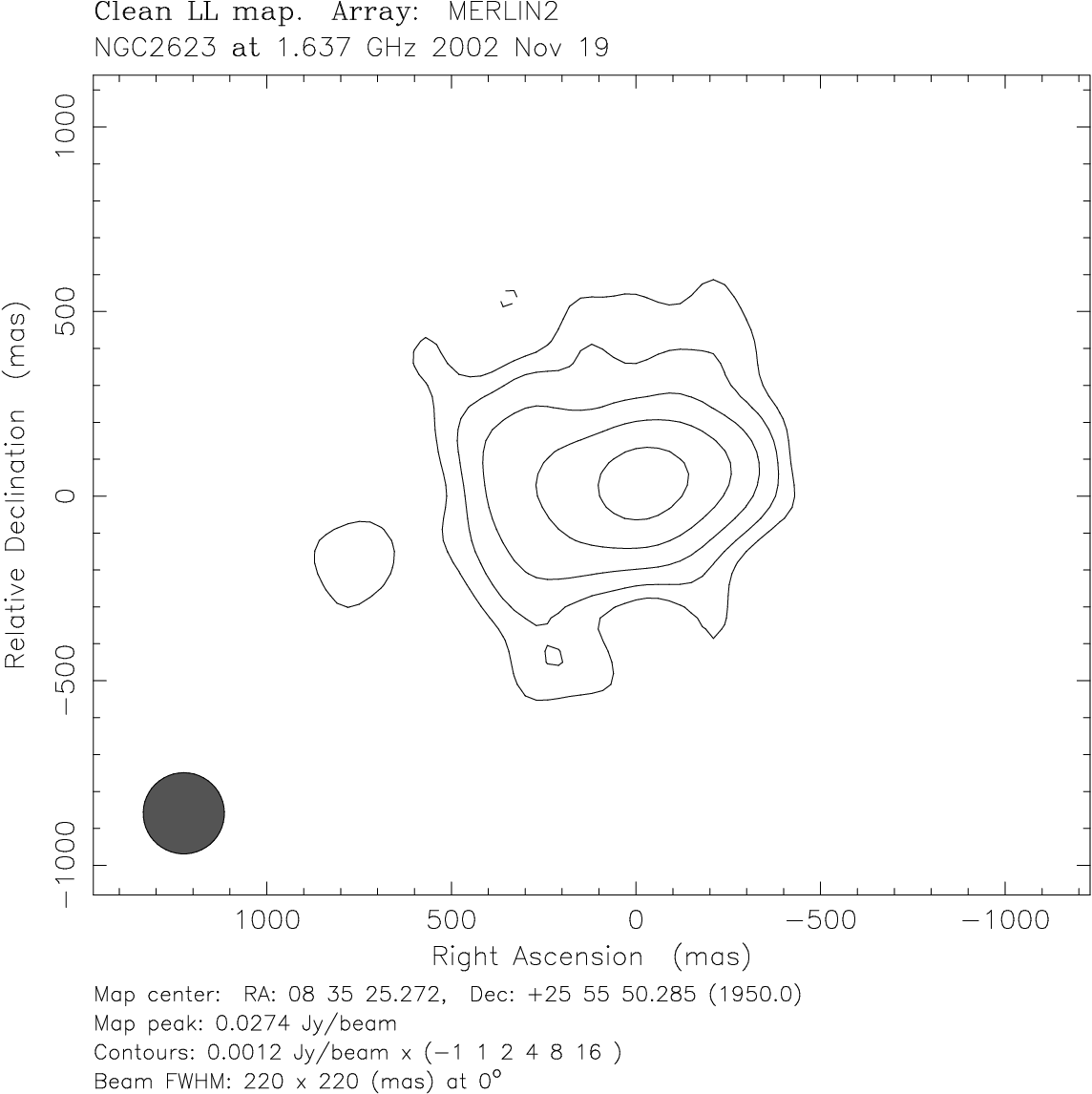}
   \includegraphics[width=7cm]{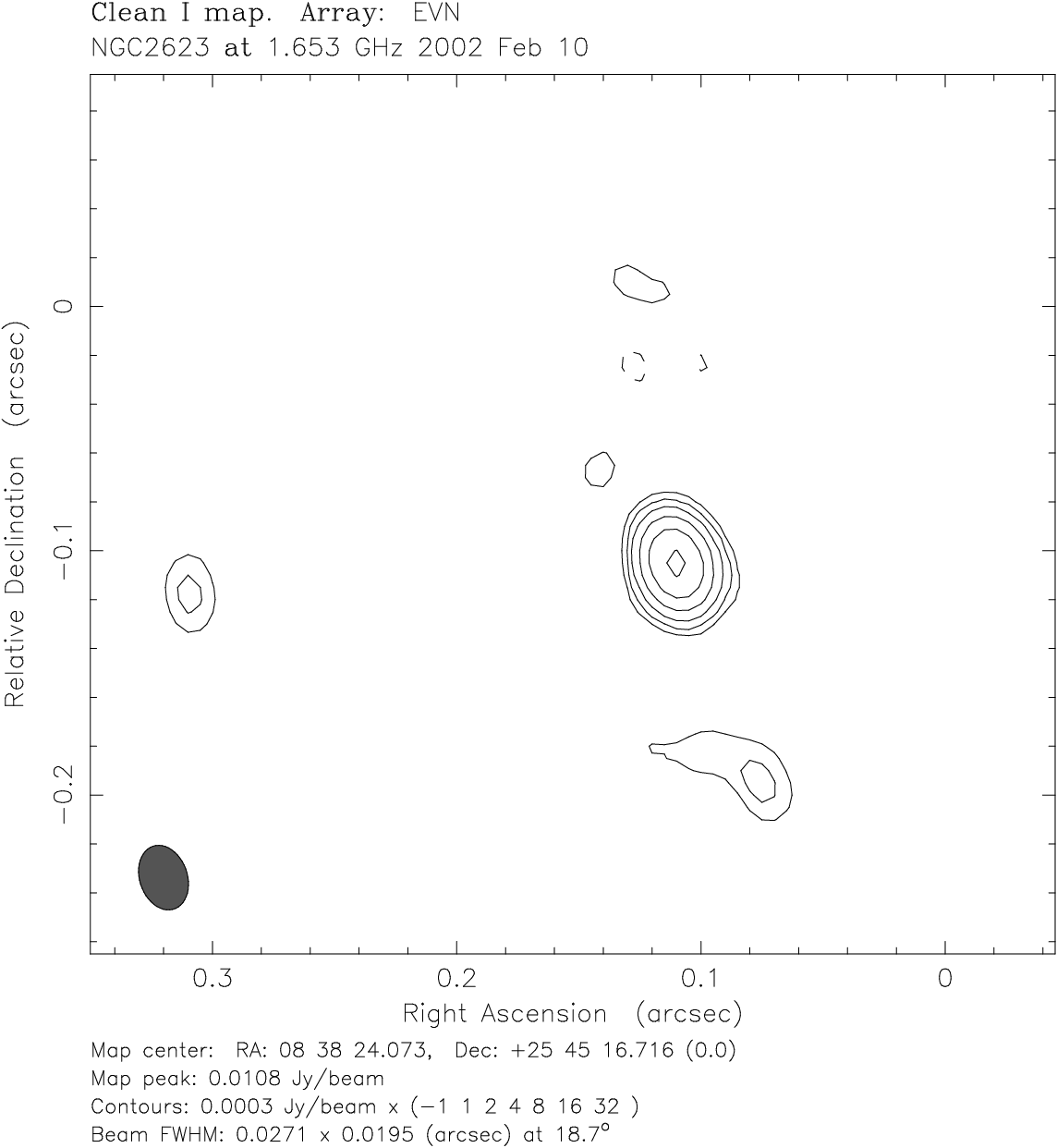} 
   \caption{Left: The MERLIN image of NGC 2623 with a restoring beam of
  220 mas. The peak is 27.4 mJy/bm and the contours are drawn at
  1,2,4,8,16,32 x 1.2 mJy/bm. Right: The EVN image of NGC 2623
  restored with a beam of 27x19 mas.  The
  peak is 10.8 mJy/bm and the contours are drawn at 1,2,4,8,16 x 3mJy/bm.
            \label{fig2623}
           }
    \end{figure*}

NGC 2623 is a LINER galaxy (at a distance d=74.138 Mpc) whose optical
appearance suggests a late state of merging. A single bright component is
seen at the centre of the galaxy at near infrared images (the nucleus
is heavily obscured at the V band). Lipari et al (\cite{lipari04})
find evidence for two starburst events, a circumnuclear burst with an
age of 1 Gyr and a compact nuclear starburst with an age of 10 Myr.

Condon et al. (\cite{condon91}) detected a well resolved structure
with the VLA at 8 GHz indicating a starburst origin for the radio
emission, while Lonsdale et al (\cite{L93}) found evidence for a high
brightness compact component embedded in complicated radio emission.

Our MERLIN image (Fig. \ref{fig2623}) shows a slightly asymmetric
structure, elongated in a roughly east-west direction, extending for
$\sim$ 350 pc. The EVN image shows at least two ($\sim$ 6x6 pc)
components, separated by 70 pc, with the brightest having luminosity
of 1.03 10$^{39}$ erg/s and a brightness temperature of 3.9 10$^{6}$K
while the weaker has a luminosity of 8.1 10$^{37}$ erg/s and a
brightness temperature of 3.9 10$^{5}$K. At least one other weak and
extended component, is detected $\sim$96 pc to the south of the
brightest component with T$_{B}= 9.6\,  10^{4}$K and a luminosity of 5.1
10$^{37}$ erg/s.

Given that the shape of the hard X-ray spectrum can only be explained
by a Compton-thick, cold, reflection dominated AGN (Maiolino et al,
\cite{mai03}) it is plausible that the brightest component in the EVN
image is the location of a weak AGN, while for the other two
components a starburst origin is plausible.

%

\section{Conclusions}

In the four ULIRGs discussed in this paper, 50\% to 70\% of the total
flux density is resolved out at parsec scale resolution. The parsec
scale structure consists of multiple components with sizes of a few to
tens of parsecs, brightness temperatures ranging from 10$^{5}$ to
10$^{7}$K and luminosities from 10$^{37}$ to 10$^{40}$ erg/s
(UGC 05101). All components are too large to be single radio supernovae
and apart from UGC 05101, their luminosities are a few thousand times
brighter than Cas A but similar to the supernova 1989J. It is
therefore possible that at least some of these components are
clustered young radio supernovae (RSN) or supernovae remnants.

It appears that at least some of the high brightness temperature
parsec scale emission is due to an AGN, in particular in UGC 05101
whose components are much more luminous than 1989J or the supernovae in
Arp 220. This plausibility needs to be explored with further
multi-frequency observations as well as X-ray observations.

The picture that seems to be emerging is that in most ULIRGs,
starbursts provide most of the energy, however a detectable AGN is
present in a fraction of those and in those cases a significant
minority of the energy (10\%--40\%) can come from AGN
accretion. Recent X-ray observations (e.g. Ptak et al (\cite{ptak03})
add further support to this picture since strong AGN-like X-ray
sources are found with CHANDRA and Beppo-SAX in a significant fraction
of ULIRGs.

\begin{acknowledgements}
The European VLBI Network is a joint facility of European, Chinese, 
South African and other radio astronomy institutes funded by their 
national research councils.
\end{acknowledgements}

\end{document}